# Nonreciprocity of spin waves in a double layer ferromagnet with Dzyaloshinskii-Moriya interactions

O.Yu. Gorobets[1,2*], Yu.I. Gorobets[1,2], I.M. Tiukavkina[1], R.S. Gerasimenko[1]

[1]*National Technical University of Ukraine «Igor Sikorsky Kyiv Polytechnic Institute», 37 Peremohy Ave., 03056 Kyiv, Ukraine*

[2]*Institute of Magnetism of NAS and MES of Ukraine, 36b Acad. Vernadskoho Blvd., 03142 Kyiv, Ukraine*

*Correspondent author e-mail: gorobets.oksana@gmail.com, National Technical University of Ukraine «Igor Sikorsky Kyiv Polytechnic Institute», 37 Peremohy Ave., 03056 Kyiv, Ukraine

*Abstract*

In this paper, boundary conditions for Landau-Lifshitz equations at the interface between two ferromagnets with different Dzyaloshinskii-Moriya interactions are derived. We calculated and verified the analytical expression for the energy flux density continuity for the spin-wave propagation through the interface between two ferromagnets with/without Dzyaloshinskii-Moriya interactions considering the boundary conditions mentioned. Analytical expressions for reflection, transmission coefficients and the nonreciprocity factor are calculated in the case of spin-wave propagation through a double layer ferromagnet with/without Dzyaloshinskii-Moriya interactions in the first/second layer. Two principally different types of nonreciprocity effects for spin waves are revealed in such a double layer system. The material parameters of a double layer ferromagnet with/without Dzyaloshinskii-Moriya interactions in the first/second layer are found for which the extremely high nonreciprocity factor (>10) is expected according to the results of calculations. The results of the paper deepen the knowledge about the spin-wave propagation control in magnonic devices.

*Introduction*

To date, ferromagnets with Dzyaloshinskii-Moriya interaction (DMI) [1,2] are the subject of intensive research in the field of magnonics due to two reasons. First, magnetic skyrmions [3–8], chiral states of domain walls [4,7,9–12] and other chiral magnetic structures [13–18] are stabilized in such materials. Moreover, the abovementioned formations are promising candidates for the role of carriers of bits of information. Second, unidirectional beams, caustics of spin waves, are formed, and the effect of nonreciprocity in the propagation of spin waves manifests itself. The last two properties allow designing unidirectional waveguides for SWs [19–22]. We expect that research in this direction will open new opportunities for SWs guidance and manipulation in ultrathin magnetic nanostructures [23–25] and will contribute to the development of beam magnonics [23].

The DMI can be created experimentally in several ways. For example, by applying a layer of heavy metal to a ferromagnetic plate [20] and by applying an external electric field [26] as an example of the topological magnon effect, namely the Aharonov-Casher effect in an external electric field [27].

Analytic expressions for spin-wave properties show that the interfacial Dzyaloshinskii-Moriya interaction leads to nonreciprocal spin-wave propagation, i.e. different properties for spin waves propagation in opposite directions [6,28] In favorable situations, it can increase the spin-wave attenuation length [28]. Besides, SWs can probe the DMI, which gives rise to topological spin textures [13].

Yet, the theoretical ideas about boundary conditions for the Landau-Lifshitz equation between ferromagnets with different magnitudes of the DMI are developed insufficiently. Due to this fact, the effective use of two- and multilayer ferromagnetic materials with Dzyaloshinskii-Moriya interaction, including magnon crystals, for creating waveguides as well as logic elements for spin waves is limited.

Thus, the paper [29] derives the most general boundary conditions for the Landau-Lifshitz equation between two ferromagnets, which are applicable also for a finite-thickness interface with a complex internal structure, i.e. with its own intrinsic magnetic anisotropy of the interface, its own Dzyaloshinskii-Moriya interaction within the interface and the arbitrary magnitude of the isotropic interchange between ferromagnets through the interface. However, the boundary conditions from this work are not applicable at the interface between ferromagnets with different values of the DMI. At the same time, the problem of boundary conditions for the magnetization vector was solved in the case of spin-wave propagation through a ferromagnet [30], in which the bulk energy density of the DMI changes abruptly when passing through a certain plane perpendicular to the spin-wave propagation direction without changing all other parameters of the ferromagnet. Therefore, these boundary conditions are not applicable to two-layer ferromagnetic materials with different parameters (saturation magnetization, magnetic anisotropy, etc.), and also do not take into account the properties of the interface itself [29].

In this paper, we obtained general boundary conditions for the Landau-Lifshitz equation at the interface between two ferromagnets with different Dzyaloshinskii-Moriya energy densities considering the complex internal structure of the interface (intrinsic magnetic anisotropy of the interface, isotropic exchange interaction, etc.). We also illustrated the application of these boundary conditions for the calculation of spin-wave propagation through a system consisting of two ferromagnetic plates with and without the DMI, respectively, and with a thickness much greater than the SW length, separated by a flat interface with a given homogeneous exchange.

*Theory and calculation*

The system consists of two semi-infinite long uniaxial ferromagnets, which are characterized by the saturation magnetizations $M_{01}$ and $M_{02}$, the exchange stiffness constants $\alpha_1(x)$, $\alpha_2(x)$ and the uniaxial magnetic anisotropy constants of ferromagnets and interface $\beta_1(x)$, $\beta_2(x)$, $\beta'(x)$ correspondingly. These ferromagnets are in contact along the plane YOZ. The external homogeneous constant magnetic field $\vec{H}_0$ is directed along the easy axis OZ.

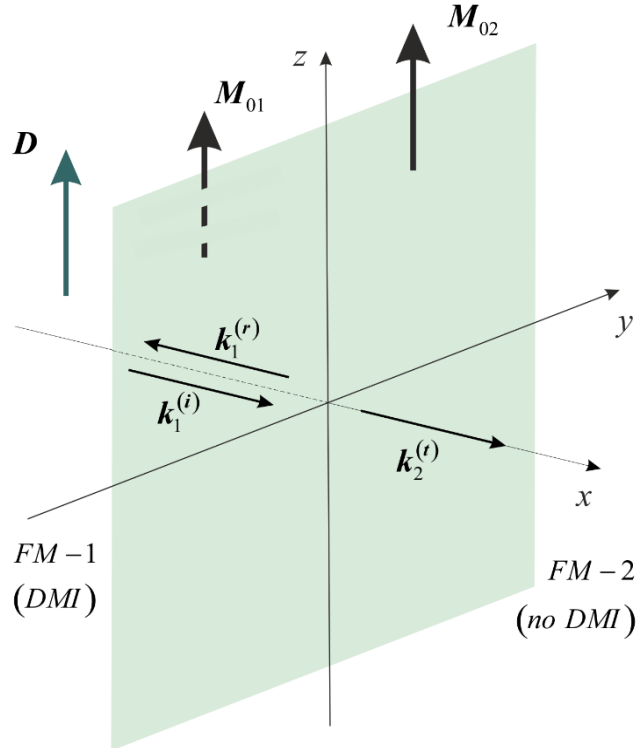

Fig.1. A schematic image of the system of two semi-infinite ferromagnetic media FM-1 and FM-2 separated by the interface of thickness $\Delta$ which is much less than the SW length. The SW is incident normally on the interface with the wave vector $\vec{k}_1^{(i)}$ and reflects with the wave vector $\vec{k}_1^{(r)}$. The transmitted SW has the wave vector $\vec{k}_1^{(t)}$. $\vec{D}$ represents the DMI vector.

The Landau-Lifshitz equations for the magnetization vector without dissipation in the first and second ferromagnets have form

$$\begin{cases} \dfrac{\partial \vec{M}_1}{\partial t} = \gamma \left[ \vec{M}_1 \times \vec{H}_{eff}^{(1)} \right], \\ \dfrac{\partial \vec{M}_2}{\partial t} = \gamma \left[ \vec{M}_2 \times \vec{H}_{eff}^{(2)} \right], \end{cases} \quad (1)$$

where $\vec{M}_1$ and $\vec{M}_2$ are the magnetization vectors in the first and second ferromagnets, respectively, $\gamma$ is the gyromagnetic ratio, $\vec{H}_{eff}^{(1)}$ and $\vec{H}_{eff}^{(2)}$ are effective magnetic fields in the first and second ferromagnets correspondingly

$$\begin{aligned} \vec{H}_{eff}^{(1)} &= -\dfrac{\delta W}{\delta \vec{M}_1}, \\ \vec{H}_{eff}^{(2)} &= -\dfrac{\delta W}{\delta \vec{M}_2}, \end{aligned} \quad (2)$$

The total energy of the system is as follows

$$W = \int_V d\vec{r} \left\{ -A(x)\vec{M}_1\vec{M}_2 + \Theta(-x)\left( \dfrac{1}{2}\alpha_1(x)\left(\dfrac{\partial \vec{M}_1}{\partial x}\right)^2 - \dfrac{1}{2}\beta_1(x)\left(\vec{M}_1\vec{n}_1\right)^2 - \vec{M}_1\vec{H}_0^{(i)} - \right.\right.$$
$$\left. -\dfrac{D_1}{M_{01}^2}\left[ \vec{e}_z\left(\vec{M}_1 \times \dfrac{\partial \vec{M}_1}{\partial x}\right) - \vec{e}_x\left(\vec{M}_1 \times \dfrac{\partial \vec{M}_1}{\partial z}\right)\right]\right) + \Theta(x)\left( \dfrac{1}{2}\alpha_2(x)\left(\dfrac{\partial \vec{M}_2}{\partial x}\right)^2 - \right.$$
$$\left.\left. -\dfrac{1}{2}\beta_2(x)\left(\vec{M}_2\vec{n}_2\right)^2 - \vec{M}_2\vec{H}_0^{(i)} - \dfrac{D_2}{M_{02}^2}\left[\vec{e}_z\left(\vec{M}_2 \times \dfrac{\partial \vec{M}_2}{\partial x}\right) - \vec{e}_x\left(\vec{M}_2 \times \dfrac{\partial \vec{M}_2}{\partial z}\right)\right]\right)\right\}, \quad (3)$$

where $A(x)$ is the parameter of uniform exchange interaction at the interface between ferromagnets. This function of the $x$ coordinate is zero within both ferromagnets and not zero within the interface, $A(x) = A\delta(x)$ where $\delta(x)$ is the Dirac delta function.

$\Theta(-x)$ and $\Theta(x)$ are the theta functions. $\alpha_1 = \lim_{x \to -\infty} \alpha_1(x)$, $\alpha_2 = \lim_{x \to \infty} \alpha_2(x)$ are the exchange stiffness constants in the first and second ferromagnets at distances much greater than the interface thickness $\Delta$. $\beta_1 = \lim_{x \to -\infty} \beta_1(x)$, $\beta_2 = \lim_{x \to \infty} \beta_2(x)$ are the uniaxial magnetic anisotropy constants in the first and second ferromagnets at distances much greater than the interface thickness $\Delta$. Vectors $\vec{n}_1$, $\vec{n}_2$ are the unit vectors directed along the axes of uniaxial magnetic anisotropy, $\vec{e}_x$, $\vec{e}_z$ are the orts of the OX and OZ axes. $D_1$, $D_2$ are Dzyaloshinskii-Moriya constants of the first and second ferromagnets.

We consider the same directions of anisotropy for both ferromagnets and the interface $\vec{n}_1 = \vec{n}_2 = \vec{n}$, here OZ is chosen parallel to the vectors $\vec{n}_1$, $\vec{n}_2$. The normal to the interface between two ferromagnets is parallel to the OX axis.

The Landau-Lifshitz equations can be integrated along the interface thickness $\Delta$ after calculating the effective fields (2) and substituting the expressions for the effective fields (2) in the Landau-Lifshitz equations. We obtain the following boundary conditions taking a limit $\Delta \to 0$ according to the method of deriving the boundary conditions for the Landau-Lifshitz equation, described in [29] [31]:

$$\begin{cases} \left[ \vec{M}_1 \times \left( \alpha_1 \frac{\partial \vec{M}_1}{\partial x} + \frac{D_1}{M_{01}^2} (\vec{e}_z \times \vec{M}_1) - A\vec{M}_2 \right) \right]\bigg|_{x=0} = 0, \\ \left[ \vec{M}_2 \times \left( \alpha_2 \frac{\partial \vec{M}_2}{\partial x} - \frac{D_2}{M_{02}^2} (\vec{e}_z \times \vec{M}_2) + A\vec{M}_1 \right) \right]\bigg|_{x=0} = 0. \end{cases} \quad (4)$$

In case the first ferromagnet has a nonzero DMI $D_1 = D$, and the second ferromagnet has no DMI $D_2 = 0$, we obtain the following boundary conditions from (4)

$$\begin{cases} \left[ \vec{M}_1 \times \left( \alpha_1 \frac{\partial \vec{M}_1}{\partial x} + \frac{D}{M_{01}^2} (\vec{e}_z \times \vec{M}_1) - A\vec{M}_2 \right) \right]\bigg|_{x=0} = 0, \\ \left[ \vec{M}_2 \times \left( \alpha_2 \frac{\partial \vec{M}_2}{\partial x} + A\vec{M}_1 \right) \right]\bigg|_{x=0} = 0. \end{cases} \quad (5)$$

If $D_1 = D$, $D_2 = 0$, then the boundary conditions (4) are applicable to describe the propagation of a SW from a ferromagnet with DMI to a ferromagnet without DMI, separated by a flat interface with a given homogeneous exchange $A$. If $D_2 = D$, $D_1 = 0$, then the boundary conditions (4) are applicable to describe the propagation of a SW from a ferromagnet without DMI to a ferromagnet with DMI, separated by a flat interface with a given homogeneous exchange $A$. In addition to the terms under boundary conditions (4), which originate from the corresponding terms in the bulk energy density of ferromagnets and the interface itself – inhomogeneous and homogeneous exchange $\alpha \frac{\partial \vec{M}_2}{\partial x}$, $A\vec{M}$ – the boundary conditions (4) can be easily generalized by supplementing them with the terms that take into account the properties of a finite thickness interface with a complex internal structure (magnetic anisotropy of the interface, DMI within the interface, etc.) and explicit expressions for which are given in [29].

The magnetization vector can be represented as $\vec{M}_1 = \vec{M}_{01} + \vec{m}_1$, where $\vec{M}_{01}$ is the saturation magnetization, $\vec{m}_1$ is the small deviation magnetization from the ground state. Similar considerations are valid for the magnetization vector $\vec{M}_{02}$ in the second ferromagnet: $\vec{M}_2 = \vec{M}_{02} + \vec{m}_2$.

The flat SW is considered further in both ferromagnets. The magnetization vectors of the incident and reflected waves can be represented as the following solution of the linearized Landau-Lifshitz equation (1) in the first ferromagnet [31]

$$m_{1x} = A_0 \cos\left(k_1^{(i)} x - \omega_1^{(i)} t + \varphi_1^{(i)}\right) + R \cos\left(-k_1^{(r)} x - \omega_1^{(r)} t + \varphi_1^{(r)}\right),$$
$$m_{1y} = A_0 \sin\left(k_1^{(i)} x - \omega_1^{(i)} t + \varphi_1^{(i)}\right) + R \sin\left(-k_1^{(r)} x - \omega_1^{(r)} t + \varphi_1^{(r)}\right),$$
(6)

where $A_0$, $R$ are the amplitudes of the incident and reflected SWs, respectively, $k_1^{(i)}$, $\varphi_1^{(i)}$, $\omega_1^{(i)}$ are the wave number, phase, and frequency of the incident spin wave, $k_1^{(r)}$, $\varphi_1^{(r)}$, $\omega_1^{(r)}$ are the wave number, phase, and frequency of the reflected spin wave, $\varphi_1^{(i)} = const$, $\varphi_1^{(r)} = const$.

The magnetization components for SWs transmitted into the second ferromagnet can be represented as the following solution of the linearized Landau-Lifshitz equation (1) in the second ferromagnet [31]

$$m_{2x} = A_2 \cos\left(k_2^{(t)} x - \omega_2^{(t)} t\right),$$
$$m_{2y} = A_2 \sin\left(k_2^{(t)} x - \omega_2^{(t)} t\right),$$
(7)

where $A_2$ is the amplitude of the SW transmitted into the second ferromagnet, $k_2^{(t)}$, $\omega_2^{(t)}$ are the wave number and frequency of the transmitted SW.

Two ferromagnets with the same parameters $M_{01} = M_{02} = M_0$, $\alpha_1 = \alpha_2 = \alpha$, $\beta_1 = \beta_2 = \beta$ are considered except that there is a nonzero DMI in the first ferromagnet, and there is no DMI in the second one.

The dispersion relation for the incident SW in the first ferromagnet with DMI has the form [28]

$$\omega_1^{(i)}\left(k_1^{(i)}\right) = -\frac{2\gamma D k_1^{(i)}}{M_0} + \gamma M_0 \sqrt{\left(\alpha\left(k_1^{(i)}\right)^2 + \frac{H_0^{(i)}}{M_0} + \beta\right)^2 + 4\pi\left(\alpha\left(k_1^{(i)}\right)^2 + \frac{H_0^{(i)}}{M_0} + \beta\right)}, \qquad (8)$$

The dispersion relation for the reflected SW in the first ferromagnet with DMI has the form [28]

$$\omega_1^{(r)}\left(k_1^{(r)}\right) = \frac{2\gamma D k_1^{(r)}}{M_0} + \gamma M_0 \sqrt{\left(\alpha\left(k_1^{(r)}\right)^2 + \frac{H_0^{(i)}}{M_0} + \beta\right)^2 + 4\pi\left(\alpha\left(k_1^{(r)}\right)^2 + \frac{H_0^{(i)}}{M_0} + \beta\right)}, \qquad (9)$$

The dispersion relation for the transmitted SW in the second ferromagnet without DMI has the form [31]:

$$\omega_2^{(t)}\left(k_2^{(t)}\right) = \gamma M_0 \sqrt{\left(\alpha\left(k_2^{(t)}\right)^2 + \frac{H_0^{(i)}}{M_0} + \beta\right)^2 + 4\pi\left(\alpha\left(k_2^{(t)}\right)^2 + \frac{H_0^{(i)}}{M_0} + \beta\right)}, \qquad (10)$$

Substituting small deviations of magnetization from the ground state it is possible to write the boundary conditions (5) in the form

$$\begin{cases} \left(\alpha \frac{\partial m_{1x}}{\partial x} - A m_{2x} + A m_{1x} - \frac{D}{M_0^2} m_{1y}\right)\bigg|_{x=0} = 0, \\ \left(\alpha \frac{\partial m_{1y}}{\partial x} - A m_{2y} + A m_{1y} + \frac{D}{M_0^2} m_{1x}\right)\bigg|_{x=0} = 0, \\ \left(\alpha \frac{\partial m_{2x}}{\partial x} + A m_{1x} - A m_{2x}\right)\bigg|_{x=0} = 0, \\ \left(\alpha \frac{\partial m_{2y}}{\partial x} + A m_{1y} - A m_{2y}\right)\bigg|_{x=0} = 0. \end{cases} \qquad (11)$$

After substituting the magnetization components (6), (7) under the boundary conditions (11), we obtained the following equations:

$$\begin{cases} \left(\alpha k_1^{(i)} - \dfrac{D}{M_0^2}\right) A_0 \cos\varphi_1^{(i)} - \left(\alpha k_1^{(r)} + \dfrac{D}{M_0^2}\right) R\cos\varphi_1^{(r)} + \\ + A\left(A_0 \sin\varphi_1^{(i)} + R\sin\varphi_1^{(r)}\right) = 0, \\ AA_2 + \left(\alpha k_1^{(i)} - \dfrac{D}{M_0^2}\right) A_0 \sin\varphi_1^{(i)} - \left(\alpha k_1^{(r)} + \dfrac{D}{M_0^2}\right) R\sin\varphi_1^{(r)} - \\ - A\left(A_0 \cos\varphi_1^{(i)} + R\cos\varphi_1^{(r)}\right) = 0, \\ \alpha k_2 A_2 + A\left(A_0 \sin\varphi_1^{(i)} + R\sin\varphi_1^{(r)}\right) = 0, \\ AA_2 - A\left(A_0 \cos\varphi_1^{(i)} + R\cos\varphi_1^{(r)}\right) = 0. \end{cases} \quad (12)$$

It follows from the boundary conditions (5) that $\omega_1^{(i)} = \omega_1^{(r)} = \omega_2^{(t)} = \omega$.

Let us denote two real positive roots of both (8) and (9) equations as $\lambda$, and $\lambda'$. Then, the dependence of $\lambda$ and $\lambda'$ on $\omega$ were obtained from the equations

$$\lambda = \dfrac{1}{2}\left[\sqrt[3]{-\dfrac{r}{2}+\sqrt{L}} + \sqrt[3]{-\dfrac{r}{2}-\sqrt{L}} - \dfrac{2}{3}Q - 4c\right]^{1/2} \quad (13)$$
$$-\dfrac{1}{2}\left[\sqrt[3]{-\dfrac{r}{2}+\sqrt{L}} + \sqrt[3]{-\dfrac{r}{2}-\sqrt{L}} - \dfrac{2}{3}Q\right]^{1/2},$$

$$\lambda' = \dfrac{1}{2}\left[\sqrt[3]{-\dfrac{r}{2}+\sqrt{L}} + \sqrt[3]{-\dfrac{r}{2}-\sqrt{L}} - \dfrac{2}{3}Q - 4c\right]^{1/2} \quad (14)$$
$$+\dfrac{1}{2}\left[\sqrt[3]{-\dfrac{r}{2}+\sqrt{L}} + \sqrt[3]{-\dfrac{r}{2}-\sqrt{L}} - \dfrac{2}{3}Q\right]^{1/2},$$

where the following notations are introduced $r = -2Q^3/27 + 8QC/3 - B^2$, $L = (r/2)^2 + (p/3)^3$, $p = -Q^2/3 - 4C$, $c = \dfrac{1}{2}\left(\sqrt[3]{-\dfrac{r}{2}+\sqrt{L}} - \sqrt[3]{\dfrac{r}{2}+\sqrt{L}} + \dfrac{1}{3}Q\right) - B\left[\sqrt[3]{-\dfrac{r}{2}+\sqrt{L}} - \sqrt[3]{\dfrac{r}{2}+\sqrt{L}} - \dfrac{2}{3}Q\right]^{-1/2}$,

$B = -4\omega D\left(\gamma\alpha^2 M_0^3\right)^{-1}$, $Q = 2\alpha^{-1}\left(H_0^{(i)}/M_0 + \beta + 2\pi\right) - 4D^2\alpha^{-2}M_0^{-4}$,

$C = \alpha^{-2}\left(H_0^{(i)}/M_0 + \beta\right)^2 + 4\pi\alpha^{-2}\left(H_0^{(i)}/M_0 + \beta\right) - \omega^2\left(\alpha\gamma M_0\right)^{-2}$.

The positive real root of the equation (10) has such form

$$\sigma = \sqrt{\dfrac{1}{\alpha}\sqrt{4\pi^2 + \left(\dfrac{\omega}{\gamma M_0}\right)^2} - \dfrac{1}{\alpha}\left(\dfrac{H_0^{(i)}}{M_0} + \beta + 2\pi\right)}. \quad (15)$$

The condition should be satisfied $\omega > \gamma M_0 \sqrt{\left(\dfrac{H_0^{(i)}}{M_0} + \beta + 2\pi\right)^2 - 4\pi^2}$.

The expression for the squared reflection coefficient $\tilde{R} = R/A_0$ has the form as follows from the system (5)

$$\tilde{R}^2 = \frac{A^2\left(\alpha k_1^{(i)} - \alpha k_2^{(t)} - \dfrac{D}{M_0^2}\right)^2 + \left(\alpha k_2^{(t)}\right)^2\left(\alpha k_1^{(i)} - \dfrac{D}{M_0^2}\right)^2}{A^2\left(\alpha k_1^{(r)} + \alpha k_2^{(t)} + \dfrac{D}{M_0^2}\right)^2 + \left(\alpha k_2^{(t)}\right)^2\left(\alpha k_1^{(r)} + \dfrac{D}{M_0^2}\right)^2}, \quad (16)$$

The expression for the squared transmission coefficient $\tilde{A}_2 = A_2/A_0$ has the form as follows from the system (5)

$$\tilde{A}_2^2 = \frac{\left(A\alpha\left(k_1^{(i)} + k_1^{(r)}\right)\right)^2}{A^2\left(\alpha k_1^{(r)} + \alpha k_2^{(t)} + \dfrac{D}{M_0^2}\right)^2 + \left(\alpha k_2^{(t)}\right)^2\left(\alpha k_1^{(r)} + \dfrac{D}{M_0^2}\right)^2}. \quad (17)$$

The SW propagation from FM -1 without DMI to FM - 2 with DMI was described at [32]. The expressions for the squared reflection coefficient $\tilde{R} = R/A_0$ and for the squared transmission coefficient $\tilde{A}_2 = A_2/A_0$ were obtained

$$\tilde{R}^2 = \frac{A^2\left(\alpha k_1^{(i)} - \alpha k_2^{(t)} + \dfrac{D}{M_0^2}\right)^2 + \left(\alpha k_2^{(t)}\right)^2\left(\alpha k_1^{(i)} + \dfrac{D}{M_0^2}\right)^2}{A^2\left(\alpha k_1^{(r)} + \alpha k_2^{(t)} - \dfrac{D}{M_0^2}\right)^2 + \left(\alpha k_2^{(t)}\right)^2\left(\alpha k_1^{(r)} - \dfrac{D}{M_0^2}\right)^2}, \quad (18)$$

$$\tilde{A}_2^2 = \frac{\left(2A\alpha k_1^{(i)}\right)^2}{A^2\left(\alpha k_1^{(r)} + \alpha k_2^{(t)} - \dfrac{D}{M_0^2}\right)^2 + \left(\alpha k_2^{(t)}\right)^2\left(\alpha k_1^{(r)} - \dfrac{D}{M_0^2}\right)^2}. \quad (19)$$

The energy density of the ferromagnet with DMI has the form [31]

$$w = F_i\left(\vec{M}_i, \frac{\partial \vec{M}_i}{\partial x}\right) - \left(\vec{M}_i \cdot \vec{H}^{(e)}\right) + \frac{1}{8\pi}\left(\vec{H}_i^{(m)}\right)^2, \quad (20)$$

where $\vec{H}^{(e)}$ is the external magnetic field, $\vec{H}_i^{(m)}$ is the static magnetic field created by the magnetization $\vec{M}_i$ of $i$ ferromagnet ($i = 1, 2$).

The function $F_i$ from $\vec{M}_i$ and $\dfrac{\partial \vec{M}_i}{\partial x}$ has the following form [31]

$$F_i\left(\vec{M}_i, \frac{\partial \vec{M}_i}{\partial x}\right) = \frac{\alpha_i}{2}\left(\frac{\partial \vec{M}_i}{\partial x}\right)^2 - \frac{\beta_i}{2}\left(\vec{M}_i \cdot \vec{n}\right) - \\ -\frac{D_i}{M_{0i}^2}\left[\vec{e}_z\left(\vec{M}_i \times \frac{\partial \vec{M}_i}{\partial x}\right) - \vec{e}_x\left(\vec{M}_i \times \frac{\partial \vec{M}_i}{\partial z}\right)\right]. \tag{21}$$

The Poynting vector in the absence of electric field [31]:

$$\Pi_k = -\frac{\partial M_i}{\partial t}\frac{\partial F_i}{\partial\left(\dfrac{\partial M_i}{\partial x_k}\right)}. \tag{22}$$

Then, under the condition of continuity of the normal component of the energy flux density on the surface of the ferromagnet $\vec{\Pi}_- \cdot \vec{e}_z = \vec{\Pi}_+ \cdot \vec{e}_z$, we get the expression

$$\alpha_2 \tilde{A}_2^2 k_2^{(t)} + \alpha_1\left[\tilde{R}^2 k_1^{(r)} - k_1^{(i)} + \tilde{R}\left(k_1^{(r)} - k_1^{(i)}\right)\cos\left(\varphi_1^{(i)} - \varphi_1^{(r)}\right)\right] = \\ = \frac{D_2}{M_{02}^2}\tilde{A}_2^2 - \frac{D_1}{M_{01}^2}\left[1 + \tilde{R}^2 + 2\tilde{R}\cos\left(\varphi_1^{(i)} - \varphi_1^{(r)}\right)\right]. \tag{23}$$

In case the first ferromagnet has a nonzero DMI $D_1 = D$, and the second ferromagnet has no DMI $D_2 = 0$, the condition has the form:

$$\alpha\left[\tilde{A}_2^2 k_2^{(t)} + \tilde{R}^2 k_1^{(r)} - k_1^{(i)} + \tilde{R}\left(k_1^{(r)} - k_1^{(i)}\right)\cos\left(\varphi_1^{(i)} - \varphi_1^{(r)}\right)\right] = \\ = -\frac{D}{M_0^2}\left[1 + \tilde{R}^2 + 2\tilde{R}\cos\left(\varphi_1^{(i)} - \varphi_1^{(r)}\right)\right], \tag{24}$$

where

$$\cos\left(\varphi_1^{(i)} - \varphi_1^{(r)}\right) = \left[ A^2 \left( \alpha k_1^{(i)} - \alpha k_2^{(t)} - \frac{D}{M_0^2} \right) \left( \alpha k_1^{(r)} + \alpha k_2^{(t)} + \frac{D}{M_0^2} \right) \right.$$

$$\left. + \left(\alpha k_2^{(t)}\right)^2 \left( \alpha k_1^{(r)} + \frac{D}{M_0^2} \right) \left( \alpha k_1^{(i)} - \frac{D}{M_0^2} \right) \right] \left[ \left( A^2 \left( \alpha k_1^{(r)} + \alpha k_2^{(t)} + \frac{D}{M_0^2} \right)^2 + \left(\alpha k_2^{(t)}\right)^2 \left( \alpha k_1^{(r)} + \frac{D}{M_0^2} \right)^2 \right) \right.$$

$$\left. \left( A^2 \left( \alpha k_1^{(i)} - \alpha k_2^{(t)} - \frac{D}{M_0^2} \right)^2 + \left(\alpha k_2^{(t)}\right)^2 \left( \alpha k_1^{(i)} - \frac{D}{M_0^2} \right)^2 \right) \right]^{-1/2}.$$

$\tilde{A}_2$ and $\tilde{R}$ satisfy the condition of continuity of the normal Poynting vector component on the interface.

*Results and Discussion*

The material parameters are represented in Table 1 to analyze the expressions (16), (17) for the SWs transmission from a ferromagnet with DMI into a ferromagnet without DMI.

Table 1. The material parameters of Permalloy (Py) [33], $Ni_{0.84}Co_{0.16}$ [34], and Yttrium Iron Garnet (YIG) [35] [36] to analyze the SWs transmission from a ferromagnet with DMI into a ferromagnet without DMI.

|  | **Values of material parameters** | | |
| --- | --- | --- | --- |
|  | Py | NiCo | YIG |
| $M_0, G$ | 860 | 710 | 140 |
| $\beta$ | 0 | $8.41 \cdot 10^{-6}$ | 0.12 |
| $\alpha, cm^2$ | $1.76 \cdot 10^{-12}$ | $2.22 \cdot 10^{-12}$ | $1.86 \cdot 10^{-11}$ |

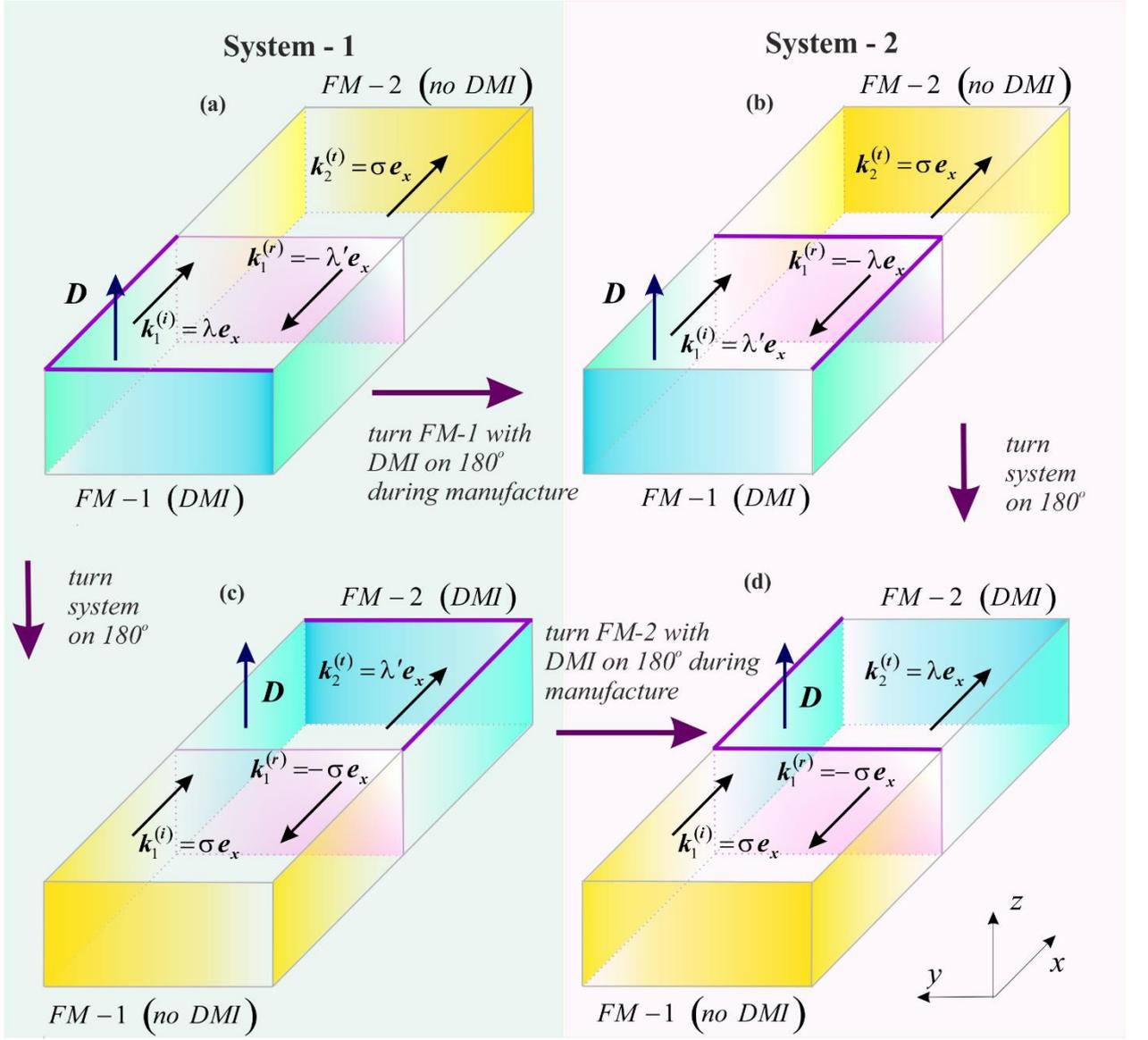

Fig.2. Nonreciprocity of spin waves in a double layer ferromagnet with Dzyaloshinskii-Moriya interactions. Cases (a) and (b) represents of SW propagation from FM -1 with DMI to FM - 2 without DMI. Cases (c) and (d) represents of SW propagation from FM -1 without DMI to FM - 2 with DMI. Purple contour shows turning FMs with DMI on 180° during manufacturing.

It is possible to make 2 types of non-equivalent double-layer systems - FM with DMI / FM without DMI. The 1$^{st}$ system is shown in Fig. 2 (a) and Fig. 2 (d). Fig. 2 (a and d) schematically shows the propagation of the spin wave in mutually opposite directions in the first system, which is equivalent to the rotation of the first system at an angle of 180° without changing the direction of propagation of the spin wave.

The nonreciprocity factor and the difference of the reflection and the transmission coefficients of spin waves, as well as in the phase differences between the incident, passing, and reflected waves will be considered below. This will be the effect of non-reciprocity of the first type, i.e., the effect of non-reciprocity in the propagation of spin waves in the 1$^{st}$ system.

The 2nd system is shown in Fig. 2 (b) and Fig. 2 (c). As can be seen from Fig. 2 (b), the rotation of FM, Fig. 2 (a), on 180º during manufacture allows getting 2$^{nd}$ system. Fig. 2 (b) and (c) schematically shows the propagation of the spin wave in mutually opposite directions in the second system, which is equivalent to the rotation of the second system at an angle of 180º without changing the direction of propagation of the spin wave. Next, we will also consider the nonreciprocity factor and the difference in the reflection coefficients, the transmission of spin waves, as well as in the phase differences between the incident, transmitted and reflected waves. This will be the effect of non-reciprocity of the 2$^{nd}$ type, i.e., the effect of non-reciprocity in the propagation of spin waves in the 2$^{nd}$ system.

Summing up, we note that there are two types of non-reciprocity effects for SW in a double-layer system FM with DMI / FM without DMI due to the possibility of manufacturing the two above-described non-equivalent double-layer systems. It is obvious that similarly there is a possibility of manufacturing 4 non-equivalent two-layer systems of type FM with DMI / FM with DMI at different non-zero values of DMI constant, and consequently 4 types of non-reciprocity effects in such systems. In this paper, for example, we consider in detail the system type FM with DMI / FM without DMI (Fig. 2).

Case 1, Fig.2. (a), was obtained by substitution of the following values of the roots of the dispersion relations (8), (9), (10) $k_1^{(i)} = \lambda$, $k_1^{(r)} = \lambda'$, $k_2^{(t)} = \sigma$ into formulas (16) and (17).

The case 2, Fig.2.(b) was obtained by substitution the roots of dispersion relations (8), (9), (10) $k_1^{(i)} = \lambda'$, $k_1^{(r)} = \lambda$, $k_2^{(t)} = \sigma$ into expressions (16) and (17).

Cases Case 3, Fig.2.(c) and Case 4, Fig.2.(d) was obtained at [32]. Although the case of SW propagation from FM-1 without DMI to FM-2 with DMI was considered, the roots of the dispersion relations have an analytical form both in formula (15) for the value of the wave vector of incident and reflected SW and for SW propagating in FM-2 as in formulas (13) or (14).

The case 3, Fig.2.(c) was obtained by substitution the wavenumbers $k_1^{(i)} = k_1^{(r)} = \sigma$, $k_2^{(t)} = \lambda$ to expressions (18) and (19). And the case 4, Fig.2. (d) was obtained by substitution the wavenumbers $k_1^{(i)} = k_1^{(r)} = \sigma$, $k_2^{(t)} = \lambda'$ to (18) and (19).

The spin waves can propagate through the two-layer systems of type FM with DMI / FM with DMI if the DMI constant is in the range of $D$, $\left[10^{-7}; 10^{-1}\right]$ $Erg/cm^2$ for FM materials NiCo, Py, and YIG because the imaginary part of wavenumbers-solutions of the equations (16)-(19) is zero for such choice of the system parameters. Modern technology allows to get DMI constant values ranges as $D$, $[0;0.012]$ $Erg/cm^2$ in insulating magnetic oxides [37], $D$, $[-1;6]$ $Erg/cm^2$ at graphene-ferromagnet interfaces [38], $D$, $[2;8]$ $Erg/cm^2$ in ultrathin films [39]. Also, giant DMI in new bilayers $D$, $[-20;30]$ $Erg/cm^2$ are theoretically predicted in [40]. The estimation of maximum values of the homogeneous exchange constant at the interface is $A \approx \dfrac{\alpha}{\Delta}$. It means that $A$ is up to $10^{-3}$ $cm$ in YIG and up to $10^{-4}$ $cm$ in NiCo and Py if we suppose the interface thickness $\Delta \approx 0.5 \cdot 10^{-3}$ $nm$. The value of $A$ as high as $10^{-2}$ $cm$ can't be achieved simply in these materials without fabricating an interface as metasurface. The values of $A \approx 10^{-2}$ $cm$ are considered in this paper as perspective for future technologies of fabrication of the two-layer ferromagnetic systems. That is why the range of $A$ is chosen as $A$, $[10^{-6};10^{2}]$ $cm$ for the analysis of the results of our model (Fig. 3 – Fig. 8).

Table 2. Nonreciprocity of spin waves in a double layer ferromagnet with Dzyaloshinskii-Moriya interactions.

| System – 1 | System – 2 |
|---|---|
| Case 1, Fig.2. (a) | Case 2, Fig.2. (b) |
| $k_1^{(i)} = \lambda$, $k_1^{(r)} = \lambda'$, $k_2^{(t)} = \sigma$. | $k_1^{(i)} = \lambda'$, $k_1^{(r)} = \lambda$, $k_2^{(t)} = \sigma$. |
| Case 3, Fig.2. (c) | Case 4, Fig.2. (d) |
| $k_1^{(i)} = k_1^{(r)} = \sigma$, $k_2^{(t)} = \lambda'$. | $k_1^{(i)} = k_1^{(r)} = \sigma$, $k_2^{(t)} = \lambda$. |

The dependences of the SW transmission coefficient $\tilde{A}_2^{\,2}$ and the SW reflection coefficient $\tilde{R}^2$ on the Dzyaloshinskii-Moriya constant $D$ are represented for Py, NiCo, and YIG (Table 1).

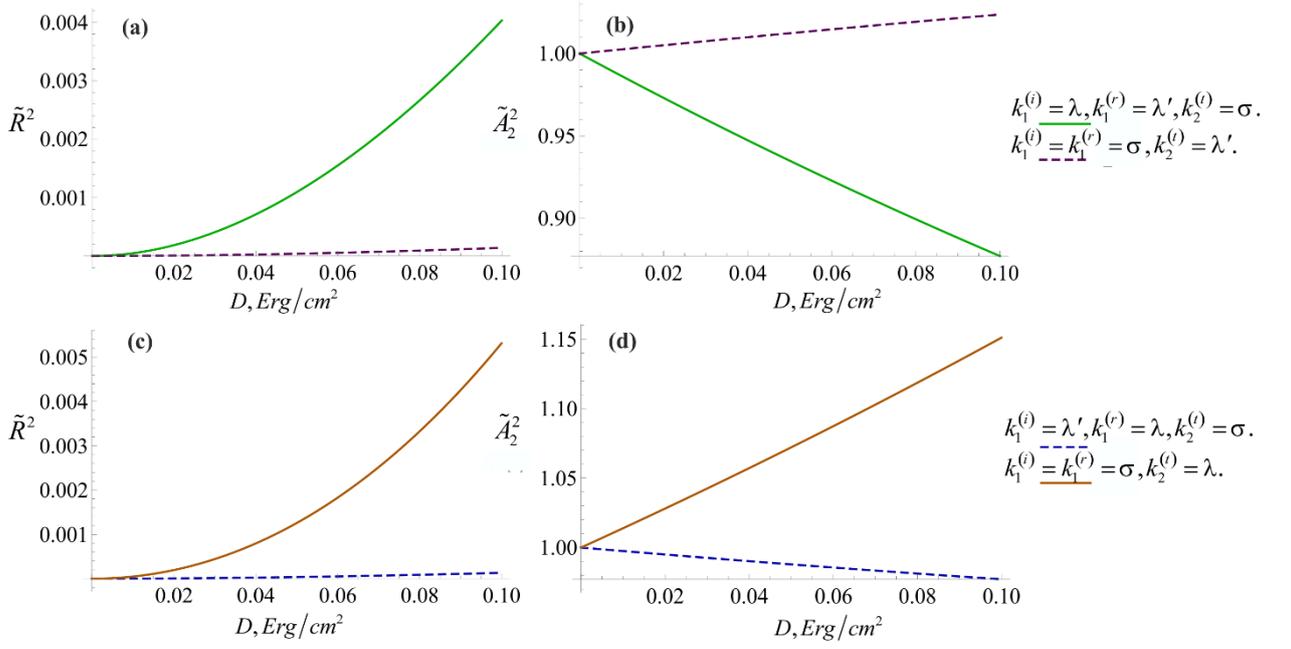

Fig.3. The dependence of the SW reflection coefficient $\tilde{R}^2$ (a, c) and the SW transmission coefficient $\tilde{A}_2^{\,2}$ (b, d) on the Dzyaloshinskii-Moriya constant $D$ for Py. Curves are plotted for different values of frequency $\omega = 90$ $GHz$. The curves represent Cases 1,3 (a), (b) and Cases 2,4 (c), (d) (Fig.2., Table 2). The parameters of uniform exchange interaction at the interface between ferromagnets are $A = 10^{-2}$ $cm$.

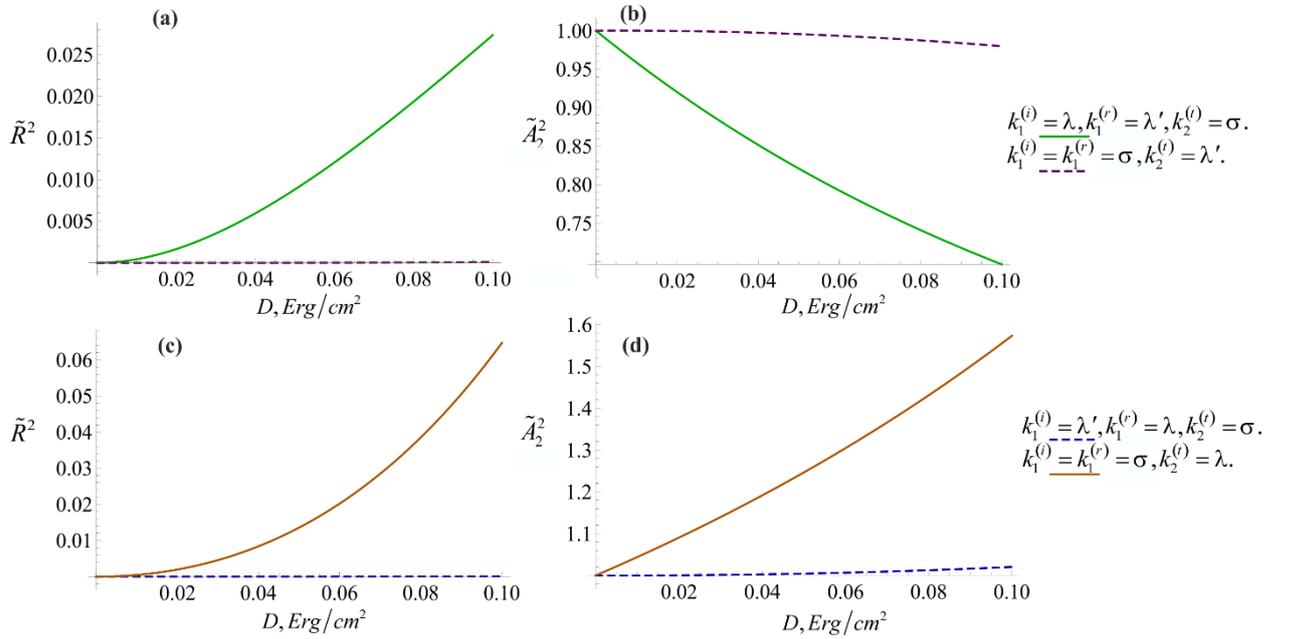

Fig.4. The dependence of the SW reflection coefficient $\tilde{R}^2$ (a, c) and the SW transmission coefficient $\tilde{A}_2^{\,2}$ (b, d) on the Dzyaloshinskii-Moriya constant $D$ for YIG. Curves are plotted for different values of frequency $\omega = 90$ $GHz$. The curves represent Cases 1,3 (a), (b) and Cases 2,4

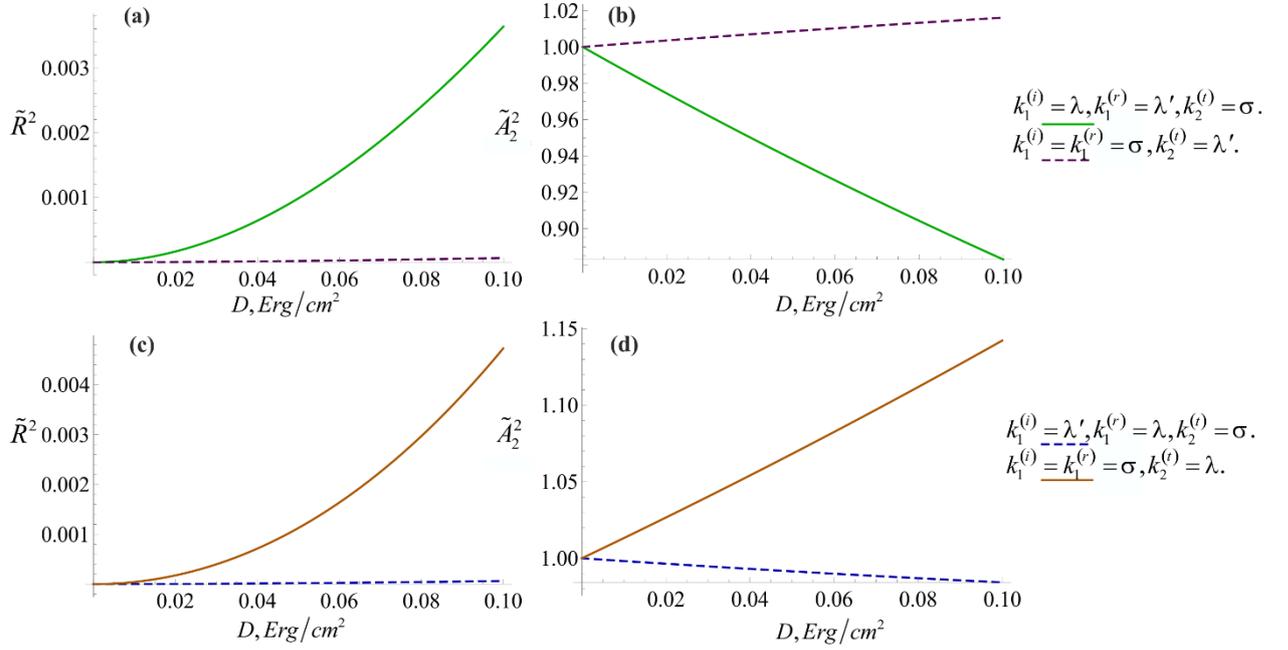

Fig.5. The dependence of the SW reflection coefficient $\tilde{R}^2$ (a, c) and the SW transmission coefficient $\tilde{A}_2^{\,2}$ (b, d) on the Dzyaloshinskii-Moriya constant $D$ for NiCo. Curves are plotted for different values of frequency $\omega = 90$ $GHz$. The curves represent Cases 1,3 (a), (b) and Cases 2,4 (c), (d) (Fig.2., Table 2). The parameters of uniform exchange interaction at the interface between ferromagnets are $A = 10^{-2}$ $cm$.

For a composite interface containing a layer of nonmagnetic material with RKKY interaction through it, the exchange decreases, as illustrated in Figures 6, 7 which leads to a decline in the intensity of transmitted SWs [41].

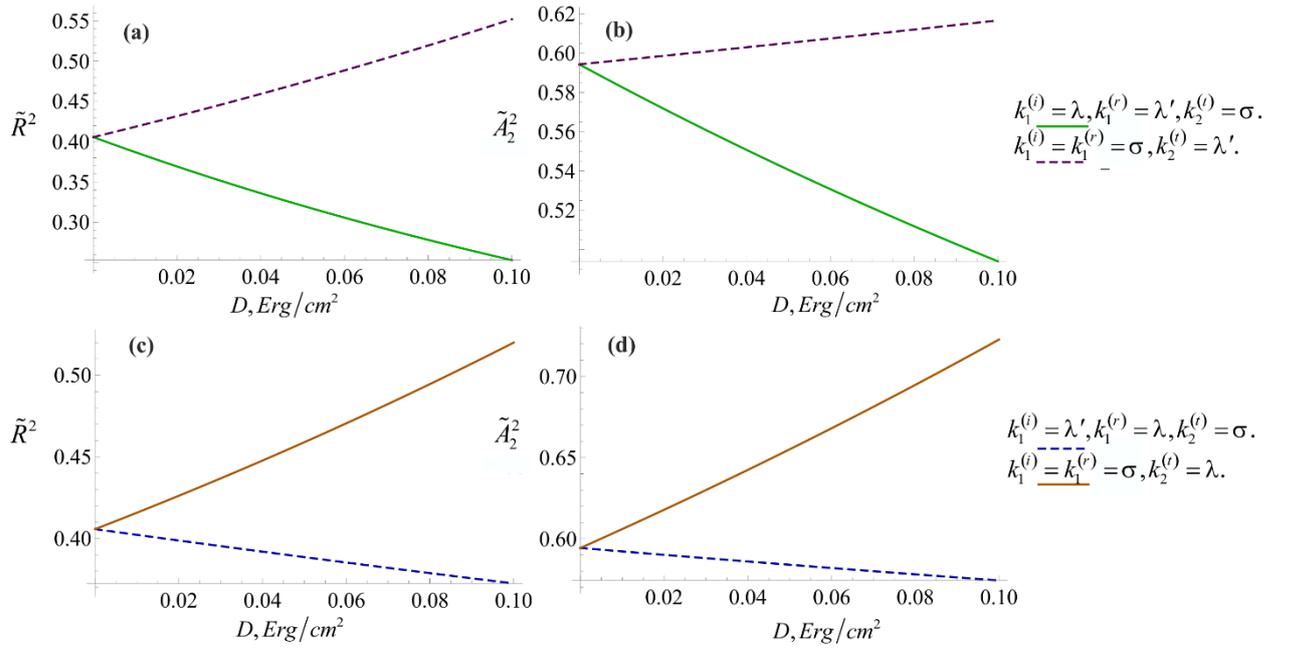

Fig.6. The dependence of the SW reflection coefficient $\tilde{R}^2$ (a, c) and the SW transmission coefficient $\tilde{A}_2^{\,2}$ (b, d) on the Dzyaloshinskii-Moriya constant $D$ for Py. Curves are plotted for different values of frequency $\omega = 90\ GHz$. The curves represent Cases 1,3 (a), (b) and Cases 2,4 (c), (d) (Fig.2., Table 2). The parameters of uniform exchange interaction at the interface between ferromagnets are $A = 10^{-6}\ cm$.

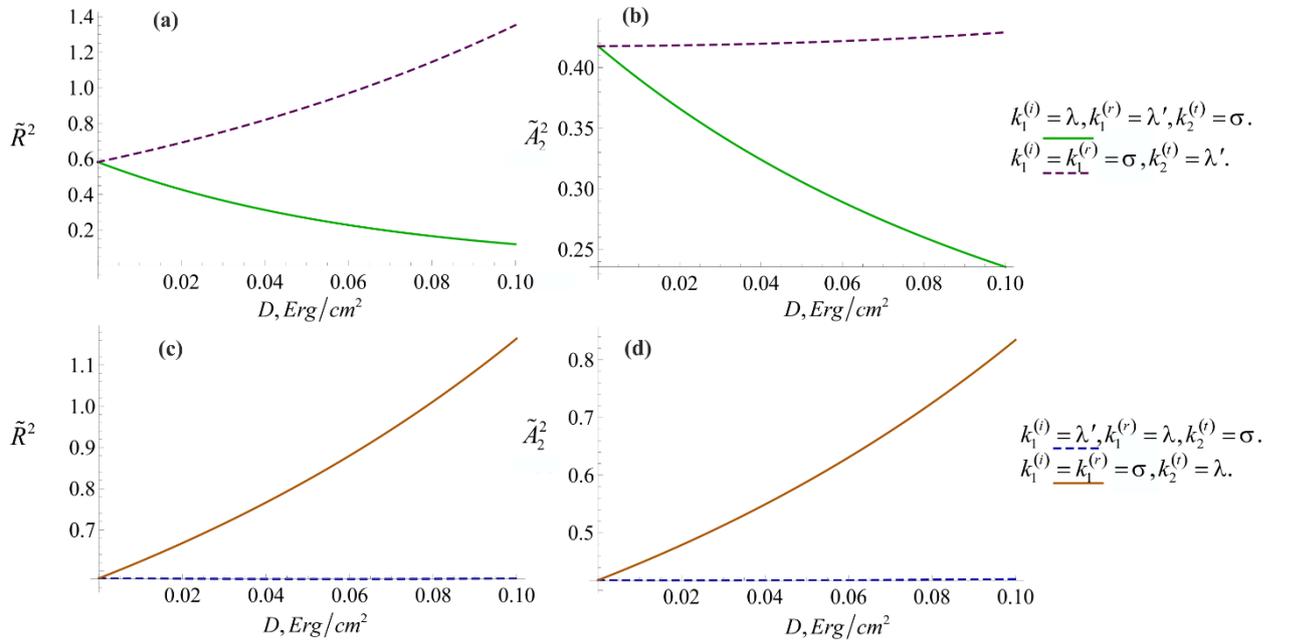

Fig.7. The dependence of the SW reflection coefficient $\tilde{R}^2$ (a, c) and the SW transmission coefficient $\tilde{A}_2^{\,2}$ (b, d) on the Dzyaloshinskii-Moriya constant $D$ for YIG. Curves are plotted for different values of frequency $\omega = 90\ GHz$. The curves represent Cases 1,3 (a), (b) and Cases 2,4

(c), (d) (Fig.2., Table 2). The parameters of uniform exchange interaction at the interface between ferromagnets are $A = 10^{-5}$ $cm$.

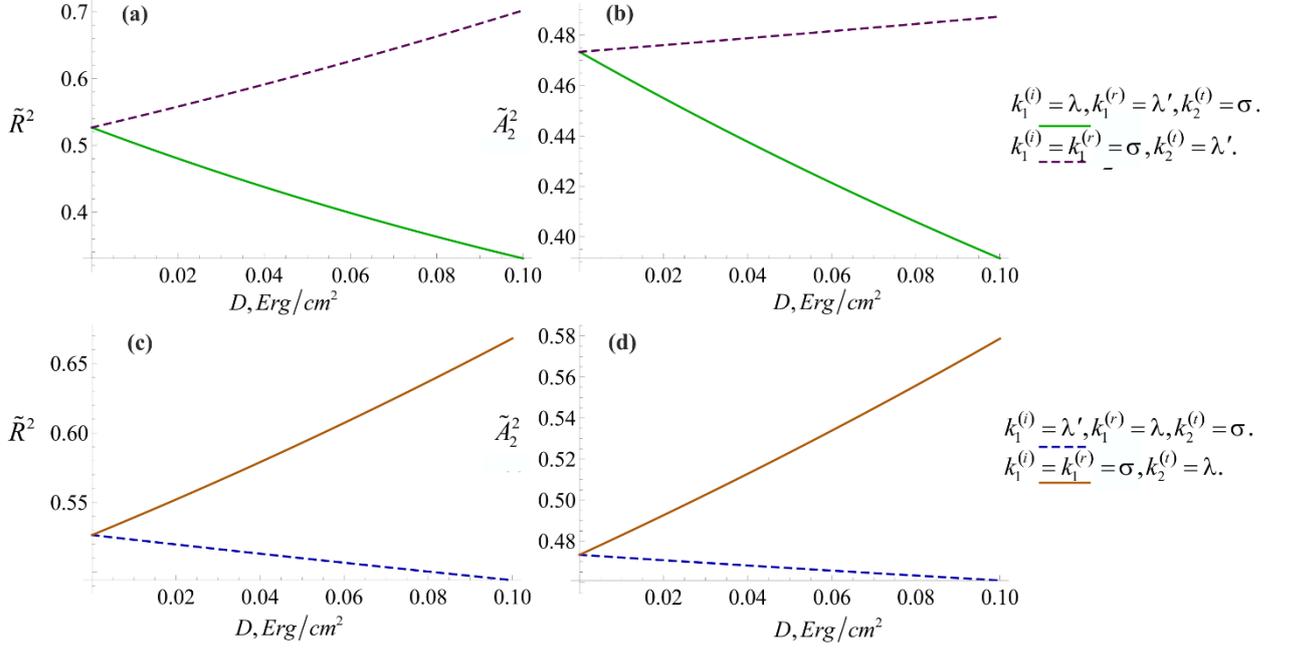

Fig.8. The dependence of the SW reflection coefficient $\tilde{R}^2$ (a, c) and the SW transmission coefficient $\tilde{A}_2^{\,2}$ (b, d) on the Dzyaloshinskii-Moriya constant $D$ for NiCo. Curves are plotted for different values of frequency $\omega = 90$ $GHz$. The curves represent Cases 1,3 (a), (b) and Cases 2,4 (c), (d) (Fig.2., Table 2). The parameters of uniform exchange interaction at the interface between ferromagnets are $A = 1.26 \cdot 10^{-6}$ $cm$.

The SW reflection and the transmission coefficient monotonically depend on the DMI in the first ferromagnet in the range $D$, $[10^{-7}; 10^{-1}]$ $Erg/cm^2$ and the SW frequency range $\omega$, $[14;120]$ $GHz$ for YIG, $\omega$, $[59;120]$ $GHz$ for Py, and $\omega$, $[10;120]$ $GHz$ for NiCo. The lower frequency limit (14 $GHz$, 59 $GHz$, and 10 $GHz$) for each of the materials is determined based on the numerical calculation of the frequency range at which the wave numbers $\lambda$, $\lambda'$, and $\sigma$ ($\lambda < \sigma < \lambda'$) for the spin wave are valid according to relations (13), (14). There is a weak dependence $\tilde{R}^2$ and $\tilde{A}_2^2$ on the frequency $\omega$ for all considered materials. $\tilde{R}^2$ and $\tilde{A}_2^2$ change by thousands of percentages in the specified frequency range.

The absolute value of the difference between the reflection coefficient of the SW propagating along the *x*-axis and the reflection coefficient propagating antiparallel to the *x*-axis increases monotonically with increasing DMI constant for all systems and materials considered in this paper. As a result, it is necessary to achieve the maximum possible value of the DMI constant in one of the ferromagnets to obtain the maximum nonreciprocity effect for SW in double-layer system (FM with DMI/FM without DMI). To analyze the maximum achievable effect of nonreciprocity in the range of the DMI constant`s values from $10^{-7}$ to $10^{-1}$ $Erg/cm^2$, we introduce the following coefficients:

$$\tilde{r} = \left( \tilde{R}^2 \big|_{D=0.1} - \tilde{R}^2 \big|_{D=0} \right) 100\%, \quad \tilde{a}_2 = \left( \tilde{A}_2^2 \big|_{D=0.1} - \tilde{A}_2^2 \big|_{D=0} \right) 100\%,$$

where $\tilde{r}$ and $\tilde{a}_2$ characterize the influence of DMI interaction on spin wave reflection and transmission. The coefficients $\Delta\tilde{r}_{13}$, $\Delta\tilde{r}_{24}$ and $\Delta\tilde{a}_{13}$, $\Delta\tilde{a}_{24}$ characterize the magnitude of the nonreciprocity effect for the reflection and the transmission coefficients of SW for two systems (indexes 1 to 4 represent the number of cases from Table 2).

$$\Delta\tilde{r}_{13} = \left[ \left| \tilde{R}^2 \big|_{D=0.1} \big|_{\substack{k_1^{(i)}=\lambda \\ k_1^{(r)}=\lambda' \\ k_2^{(t)}=\sigma}} - \tilde{R}^2 \big|_{D=0.1} \big|_{\substack{k_1^{(i)}=\sigma \\ k_1^{(r)}=\sigma \\ k_2^{(t)}=\lambda'}} \right| \right] 100\%,$$

$$\Delta\tilde{r}_{24} = \left[ \left| \tilde{R}^2 \big|_{D=0.1} \big|_{\substack{k_1^{(i)}=\lambda' \\ k_1^{(r)}=\lambda \\ k_2^{(t)}=\sigma}} - \tilde{R}^2 \big|_{D=0.1} \big|_{\substack{k_1^{(i)}=\sigma \\ k_1^{(r)}=\sigma \\ k_2^{(t)}=\lambda}} \right| \right] 100\%,$$

$$\Delta\tilde{a}_{13} = \left[ \left| \tilde{A}_2^2 \big|_{D=0.1} \big|_{\substack{k_1^{(i)}=\lambda \\ k_1^{(r)}=\lambda' \\ k_2^{(t)}=\sigma}} - \tilde{A}_2^2 \big|_{D=0.1} \big|_{\substack{k_1^{(i)}=\sigma \\ k_1^{(r)}=\sigma \\ k_2^{(t)}=\lambda'}} \right| \right] 100\%,$$

$$\Delta\tilde{a}_{24} = \left[ \left| \tilde{A}_2^2 \big|_{D=0.1} \big|_{\substack{k_1^{(i)}=\lambda' \\ k_1^{(r)}=\lambda \\ k_2^{(t)}=\sigma}} - \tilde{A}_2^2 \big|_{D=0.1} \big|_{\substack{k_1^{(i)}=\sigma \\ k_1^{(r)}=\sigma \\ k_2^{(t)}=\lambda}} \right| \right] 100\%.$$

To characterize the magnitude of the nonreciprocity effect, we introduce the nonreciprocity factor as the ratio of the SW amplitude at opposite the wave vector by analogy with the approach in the work [42].

$$NR_{13}^{(r)} = \left| \max\left\{ \tilde{R}^2 \big|_{\substack{k_1^{(i)}=\lambda \\ k_1^{(r)}=\lambda' \\ k_2^{(t)}=\sigma}}, \tilde{R}^2 \big|_{\substack{k_1^{(i)}=\sigma \\ k_1^{(r)}=\sigma \\ k_2^{(t)}=\lambda'}} \right\} \Big/ \min\left\{ \tilde{R}^2 \big|_{\substack{k_1^{(i)}=\lambda \\ k_1^{(r)}=\lambda' \\ k_2^{(t)}=\sigma}}, \tilde{R}^2 \big|_{\substack{k_1^{(i)}=\sigma \\ k_1^{(r)}=\sigma \\ k_2^{(t)}=\lambda'}} \right\} \right|_{D=0.1},$$

$$NR_{24}^{(r)} = \left| \max\left\{ \tilde{R}^2 \big|_{\substack{k_1^{(i)}=\lambda' \\ k_1^{(r)}=\lambda \\ k_2^{(t)}=\sigma}}, \tilde{R}^2 \big|_{\substack{k_1^{(i)}=\sigma \\ k_1^{(r)}=\sigma \\ k_2^{(t)}=\lambda}} \right\} \Big/ \min\left\{ \tilde{R}^2 \big|_{\substack{k_1^{(i)}=\lambda \\ k_1^{(r)}=\lambda' \\ k_2^{(t)}=\sigma}}, \tilde{R}^2 \big|_{\substack{k_1^{(i)}=\sigma \\ k_1^{(r)}=\sigma \\ k_2^{(t)}=\lambda'}} \right\} \right|_{D=0.1},$$

$$NR_{13}^{(t)} = \left| \max\left\{ \tilde{A}_2^2 \big|_{\substack{k_1^{(i)}=\lambda \\ k_1^{(r)}=\lambda' \\ k_2^{(t)}=\sigma}}, \tilde{A}_2^2 \big|_{\substack{k_1^{(i)}=\sigma \\ k_1^{(r)}=\sigma \\ k_2^{(t)}=\lambda'}} \right\} \Big/ \min\left\{ \tilde{A}_2^2 \big|_{\substack{k_1^{(i)}=\lambda \\ k_1^{(r)}=\lambda' \\ k_2^{(t)}=\sigma}}, \tilde{A}_2^2 \big|_{\substack{k_1^{(i)}=\sigma \\ k_1^{(r)}=\sigma \\ k_2^{(t)}=\lambda'}} \right\} \right|_{D=0.1},$$

$$NR_{24}^{(t)} = \left| \max\left\{ \tilde{A}_2^2 \big|_{\substack{k_1^{(i)}=\lambda' \\ k_1^{(r)}=\lambda \\ k_2^{(t)}=\sigma}}, \tilde{A}_2^2 \big|_{\substack{k_1^{(i)}=\sigma \\ k_1^{(r)}=\sigma \\ k_2^{(t)}=\lambda}} \right\} \Big/ \min\left\{ \tilde{A}_2^2 \big|_{\substack{k_1^{(i)}=\lambda' \\ k_1^{(r)}=\lambda \\ k_2^{(t)}=\sigma}}, \tilde{A}_2^2 \big|_{\substack{k_1^{(i)}=\sigma \\ k_1^{(r)}=\sigma \\ k_2^{(t)}=\lambda}} \right\} \right|_{D=0.1}.$$

Table 3. The coefficients $\tilde{r}$, $\tilde{a}_2$, $\Delta\tilde{r}_{13}$, $\Delta\tilde{r}_{24}$, $\Delta\tilde{a}_{13}$, $\Delta\tilde{a}_{24}$, $NR_{13}^{(r)}$, $NR_{24}^{(r)}$, $NR_{13}^{(t)}$, and $NR_{24}^{(t)}$ are represented for Py, NiCo, YIG in Table 1.

| $A$, cm | Py | | NiCo | | YIG | |
|---|---|---|---|---|---|---|
| | FM with DMI => FM without DMI $k_1^{(i)} = \lambda$, $k_1^{(r)} = \lambda'$, $k_2^{(t)} = \sigma$. | FM without DMI => FM with DMI $k_1^{(i)} = \sigma$, $k_1^{(r)} = \sigma$, $k_2^{(t)} = \lambda'$. | FM with DMI => FM without DMI $k_1^{(i)} = \lambda$, $k_1^{(r)} = \lambda'$, $k_2^{(t)} = \sigma$. | FM without DMI => FM with DMI $k_1^{(i)} = \sigma$, $k_1^{(r)} = \sigma$, $k_2^{(t)} = \lambda'$. | FM with DMI => FM without DMI $k_1^{(i)} = \lambda$, $k_1^{(r)} = \lambda'$, $k_2^{(t)} = \sigma$. | FM without DMI => FM with DMI $k_1^{(i)} = \sigma$, $k_1^{(r)} = \sigma$, $k_2^{(t)} = \lambda'$. |
| $10^{-2}$ | $\tilde{r} = -18\%$, $\tilde{a}_2 = -115\%$ | $\tilde{r} = -19\%$, $\tilde{a}_2 = 70\%$ | $\tilde{r} = -1\%$, $\tilde{a}_2 = -38\%$ | $\tilde{r} = -2\%$, $\tilde{a}_2 = 25\%$ | $\tilde{r} = 2\%$, $\tilde{a}_2 = -43\%$ | $\tilde{r} = -0.4\%$, $\tilde{a}_2 = 10\%$ |
| | $\Delta\tilde{r}_{13} = 0.3\%$, $\Delta\tilde{a}_{13} = 12\%$, $NR_{13}^{(r)} = 29.4$, $NR_{13}^{(t)} = 1.1$ | | $\Delta\tilde{r}_{13} = 0.4\%$, $\Delta\tilde{a}_{13} = 13\%$, $NR_{13}^{(r)} = 56.2$, $NR_{13}^{(t)} = 1.2$ | | $\Delta\tilde{r}_{13} = 3\%$, $\Delta\tilde{a}_{13} = 28\%$, $NR_{13}^{(r)} = 259.8$, $NR_{13}^{(t)} = 1.4$ | |
| $10^{-5}$ | $\tilde{r} = -19\%$, $\tilde{a}_2 = -112\%$ | $\tilde{r} = -19\%$, $\tilde{a}_2 = 69\%$ | $\tilde{r} = -2\%$, $\tilde{a}_2 = -37\%$ | $\tilde{r} = -1\%$, $\tilde{a}_2 = 25\%$ | $\tilde{r} = -49\%$, $\tilde{a}_2 = -20\%$ | $\tilde{r} = 74\%$, $\tilde{a}_2 = 9\%$ |
| | $\Delta\tilde{r}_{13} = 0.3\%$, $\Delta\tilde{a}_{13} = 12\%$, $NR_{13}^{(r)} = 1.3$, $NR_{13}^{(t)} = 1.1$ | | $\Delta\tilde{r}_{13} = 0.8\%$, $\Delta\tilde{a}_{13} = 13\%$, $NR_{13}^{(r)} = 1.5$, $NR_{13}^{(t)} = 1.2$ | | $\Delta\tilde{r}_{13} = 123\%$, $\Delta\tilde{a}_{13} = 19\%$, $NR_{13}^{(r)} = 11.3$, $NR_{13}^{(t)} = 1.8$ | |
| $10^{-6}$ | $\tilde{r} = -40\%$, $\tilde{a}_2 = -25\%$ | $\tilde{r} = -9\%$, $\tilde{a}_2 = 41\%$ | $\tilde{r} = -30\%$, $\tilde{a}_2 = -10\%$ | $\tilde{r} = 16\%$, $\tilde{a}_2 = 14\%$ | $\tilde{r} = -83\%$, $\tilde{a}_2 = -0.4\%$ | $\tilde{r} = 139\%$, $\tilde{a}_2 = 0.2\%$ |
| | $\Delta\tilde{r}_{13} = 31\%$, $\Delta\tilde{a}_{13} = 9\%$, $NR_{13}^{(r)} = 1.9$, $NR_{13}^{(t)} = 1.2$ | | $\Delta\tilde{r}_{13} = 46\%$, $\Delta\tilde{a}_{13} = 8\%$, $NR_{13}^{(r)} = 2.2$, $NR_{13}^{(t)} = 1.3$ | | $\Delta\tilde{r}_{13} = 222\%$, $\Delta\tilde{a}_{13} = 0.4\%$, $NR_{13}^{(r)} = 14.3$, $NR_{13}^{(t)} = 2.1$ | |
| | FM with DMI => FM without DMI $k_1^{(i)} = \lambda'$, $k_1^{(r)} = \lambda$, $k_2^{(t)} = \sigma$. | FM without DMI => FM with DMI $k_1^{(i)} = \sigma$, $k_1^{(r)} = \sigma$, $k_2^{(t)} = \lambda$. | FM with DMI => FM without DMI $k_1^{(i)} = \lambda'$, $k_1^{(r)} = \lambda$, $k_2^{(t)} = \sigma$. | FM without DMI => FM with DMI $k_1^{(i)} = \sigma$, $k_1^{(r)} = \sigma$, $k_2^{(t)} = \lambda$. | FM with DMI => FM without DMI $k_1^{(i)} = \lambda'$, $k_1^{(r)} = \lambda$, $k_2^{(t)} = \sigma$. | FM without DMI => FM with DMI $k_1^{(i)} = \sigma$, $k_1^{(r)} = \sigma$, $k_2^{(t)} = \lambda$. |
| $10^{-2}$ | $\tilde{r} = -19\%$, $\tilde{a}_2 = -107\%$ | $\tilde{r} = -18\%$, $\tilde{a}_2 = 80\%$ | $\tilde{r} = -2\%$, $\tilde{a}_2 = -28\%$ | $\tilde{r} = -1\%$, $\tilde{a}_2 = 38\%$ | $\tilde{r} = -0.4\%$, $\tilde{a}_2 = -10.5\%$ | $\tilde{r} = 6\%$, $\tilde{a}_2 = 69\%$ |
| | $\Delta\tilde{r}_{24} = 0.3\%$, $\Delta\tilde{a}_{24} = 14\%$, $NR_{24}^{(r)} = 38.6$, $NR_{24}^{(t)} = 1.1$ | | $\Delta\tilde{r}_{24} = 0.5\%$, $\Delta\tilde{a}_{24} = 16\%$, $NR_{24}^{(r)} = 75.7$, $NR_{24}^{(t)} = 1.2$ | | $\Delta\tilde{r}_{24} = 6.5\%$, $\Delta\tilde{a}_{24} = 55\%$, $NR_{24}^{(r)} = 628.3$, $NR_{24}^{(t)} = 1.5$ | |
| $10^{-5}$ | $\tilde{r} = -19\%$, $\tilde{a}_2 = -103\%$ | $\tilde{r} = -19\%$, $\tilde{a}_2 = 80\%$ | $\tilde{r} = -2\%$, $\tilde{a}_2 = -27\%$ | $\tilde{r} = -1\%$, $\tilde{a}_2 = 38\%$ | $\tilde{r} = -3\%$, $\tilde{a}_2 = -2\%$ | $\tilde{r} = 55\%$, $\tilde{a}_2 = 49\%$ |
| | $\Delta\tilde{r}_{24} = 0.6\%$, $\Delta\tilde{a}_{24} = 14\%$ | | $\Delta\tilde{r}_{24} = 1\%$, $\Delta\tilde{a}_{24} = 16\%$, $NR_{24}^{(r)} = 1.5$, $NR_{24}^{(t)} = 1.2$ | | $\Delta\tilde{r}_{24} = 58\%$, $\Delta\tilde{a}_{24} = 41\%$, $NR_{24}^{(r)} = 2.0$, $NR_{24}^{(t)} = 1.9$ | |
| | $NR_{24}^{(r)} = 1.6$, $NR_{24}^{(t)} = 1.1$ | | | | | |
| $10^{-6}$ | $\tilde{r} = -27\%$, $\tilde{a}_2 = -19\%$ | $\tilde{r} = -11\%$, $\tilde{a}_2 = 49\%$ | $\tilde{r} = -10\%$, $\tilde{a}_2 = -4\%$ | $\tilde{r} = 12\%$, $\tilde{a}_2 = 21\%$ | $\tilde{r} = -1\%$, $\tilde{a}_2 = -0.01\%$ | $\tilde{r} = 139\%$, $\tilde{a}_2 = 1\%$ |
| | $\Delta\tilde{r}_{24} = 15\%$, $\Delta\tilde{a}_{24} = 11\%$, $NR_{24}^{(r)} = 1.3$, $NR_{24}^{(t)} = 1.2$ | | $\Delta\tilde{r}_{24} = 22\%$, $\Delta\tilde{a}_{24} = 10\%$, $NR_{24}^{(r)} = 1.4$, $NR_{24}^{(t)} = 1.3$ | | $\Delta\tilde{r}_{24} = 140\%$, $\Delta\tilde{a}_{24} = 1\%$, $NR_{24}^{(r)} = 2.4$, $NR_{24}^{(t)} = 2.5$ | |

The table 3 shows that the maximum difference of the SW transmission coefficients $\Delta a$ and $\Delta\tilde{a}$ is achieved at large values of the homogeneous exchange constant at the interface, for example, at $A = 10^{-2}$ cm and for YIG is more than 50%. The nonreciprocity factor for the transmitted spin wave $NR^{(t)}$ is calculated in the range 1.1 – 2.5. The nonreciprocity factor for the transmitted spin wave (in the majority of cases) increases with decrease of the homogeneous exchange constant due to decrease of the minimum transmission coefficient in the denominator of expression for $NR^{(t)}$ while the maximum difference of the SW transmission coefficients $\Delta a$ and $\Delta\tilde{a}$ manifests the contrary trend.

On the contrary, at large values of the homogeneous exchange constant $A = 10^{-2}$ cm, the maximum difference of the SW reflection coefficients $\Delta r$ and $\Delta\tilde{r}$ is small (for most materials less than 1%, for YIG about 6%). Large maximum difference of the SW reflection coefficients $\Delta r$ and $\Delta\tilde{r}$ is achieved at small values of the homogeneous exchange constant at the interface, for example, at $A = 10^{-5} - 10^{-6}$ cm (more than 200% for YIG). The nonreciprocity factor for the reflected spin wave increases up to the values >600 with increase of the homogeneous exchange constant ($A = 10^{-2}$ cm) due to significant decrease of the minimum reflection coefficient in the denominator of expression for $NR^{(r)}$ while the maximum difference of the SW reflection coefficients $\Delta r$ and $\Delta\tilde{r}$ manifests the contrary trend. The nonreciprocity factor for the reflected spin wave is in the range 1.1 – 14.3 for $A = \left[10^{-6}, 10^{-5}\right]$ cm.

The condition of continuity of energy flux density at the boundary of two ferromagnets (23) derived in this paper describes the relationship between the reflection coefficients, the transmission coefficient, wave numbers and such system parameters as the DMI parameter, the exchange stiffness constants, and the inhomogeneous exchange parameter. In this case, the condition of continuity of energy flux density (23) coincides with the well-known corresponding condition $\tilde{R}^2 + \tilde{A}_2^2 = 1$ in the absence of DMI in both identical ferromagnets. This expression prohibits the values of the reflection and transmission coefficients greater than 1. An important conclusion of condition (23) is that the values of the reflection and transmission coefficients can be greater than one when the SW propagation through a double-layer system of ferromagnets with all the same physical parameters except the DMI constant.

$\tilde{A}_2^2 > 1$ from formula (19) can be obtained in the case of the SW propagation from FM without DMI to FM with DM

$$\frac{D^2}{M_0^4}\left(A^2 + \left(\alpha k_2^{(t)}\right)^2\right) + \left(\alpha k_1^{(r)}\right)^2\left(A^2 + \left(\alpha k_2^{(t)}\right)^2\right) - 2\alpha k_1^{(r)}\frac{D}{M_0^2}\left(A^2 + \left(\alpha k_2^{(t)}\right)^2\right)$$
$$+ 2A^2\alpha^2 k_2^{(t)} k_1^{(r)} + 2\frac{D}{M_0^2}A^2\alpha k_2^{(t)} + A^2\left(\alpha k_2^{(t)}\right)^2 - 4\left(A\alpha k_1^{(i)}\right)^2 < 0,$$

$\tilde{A}_2^2 > 1$ from formula (17) can be obtained in the case of the SW propagation from FM with DMI to FM without DM

$$\frac{D^2}{M_0^4}\left(A^2 + \left(\alpha k_2^{(t)}\right)^2\right) + 2\alpha \frac{D}{M_0^2}\left(A^2\left(k_1^{(r)} + k_2^{(t)}\right) + k_1^{(r)}\left(\alpha k_2^{(t)}\right)^2\right)$$
$$\alpha^2 A^2\left(k_1^{(r)} + k_2^{(t)}\right)^2 + \left(\alpha^2 k_2^{(t)} k_1^{(r)}\right)^2 - A^2\alpha^2\left(k_1^{(i)} + k_1^{(r)}\right)^2 < 0$$

As can be seen from Table 3, a double-layer system (FM with DMI/FM without DMI) can be a functional element (amplifier) to increase the amplitude of the SW passing into the second ferromagnet by 1.6 times, and the amplitude of the reflected SW by 1.4 times comparable to the amplitude of the incident SW.

Table 4. The results of the numerical calculation of the DMI constant's range for the transmission coefficient over 1, the exchange constant is $A = 10^{-2}$ cm.

|  | $\omega$, GHz | Case 2, Fig.2. (b) $k_1^{(i)} = \lambda'$, $k_1^{(r)} = \lambda$, $k_2^{(t)} = \sigma$. | Case 4, Fig.2. (d) $k_1^{(i)} = k_1^{(r)} = \sigma$, $k_2^{(t)} = \lambda$. |
|---|---|---|---|
| YIG | 90 | $0.013 < D < 0.500$ | $4.30 \cdot 10^{-6} < D < 0.51$ |
| YIG | 105 | $0.011 < D < 0.540$ | $3.97 \cdot 10^{-6} < D < 0.53$ |
| YIG | 120 | $0.009 < D < 0.570$ | $3.73 \cdot 10^{-6} < D < 0.58$ |
|  |  | Case 3, Fig.2. (c) $k_1^{(i)} = k_1^{(r)} = \sigma$, $k_2^{(t)} = \lambda'$.. | Case 4, Fig.2. (d) $k_1^{(i)} = k_1^{(r)} = \sigma$, $k_2^{(t)} = \lambda$. |
| Py | 90 | $3.22 \cdot 10^{-5} < D < 1.01$ | $1.83 \cdot 10^{-5} < D < 2.8$ |
| Py | 105 | $3.22 \cdot 10^{-5} < D < 0.91$ | $1.7 \cdot 10^{-5} < D < 2.95$ |
| Py | 120 | $3.29 \cdot 10^{-5} < D < 0.83$ | $1.61 \cdot 10^{-5} < D < 3.1$ |
| NiCo | 90 | $2.55 \cdot 10^{-5} < D < 0.68$ | $1.24 \cdot 10^{-5} < D < 2.3$ |
| NiCo | 105 | $2.47 \cdot 10^{-5} < D < 0.6$ | $1.2 \cdot 10^{-5} < D < 2.42$ |
| NiCo | 120 | $2.61 \cdot 10^{-5} < D < 0.54$ | $1.13 \cdot 10^{-5} < D < 2.54$ |

*Conclusions*

The boundary conditions for the Landau-Lifshitz equation derived in this paper are applicable to describe the problems of SW propagation through a system of two ferromagnets with different values of the DMI constants $D_1$, $D_2$ and describe the relationship between the dynamic components of the SW magnetization depending on the values of the DMI constant $D_1$, $D_2$ and the value of the homogeneous exchange constant $A$.

The case of SW propagation in a double-layer system (FM-1 with DMI / FM-2 without DMI) under the same values of other parameters of both ferromagnets (exchange constant, uniaxial magnetic anisotropy constant) is analyzed on the basis in our analytical model. The reflection and transmission coefficients of the SW monotonically depend on the Dzyaloshinskii-Moriya interaction constant in the range of constant values $D$ from $10^{-7}$ to $10^{-1}$ $Erg/cm^2$ and in the range of SW frequencies for which the wave numbers are real according to the dispersion relation. The dependence of the reflection and transmission of the spin wave on the frequency $\omega$ in the specified frequency range for all considered materials is very weak, the changes occur at the level of thousandths of a percent.

The double-layer system (FM-1 with DMI / FM-2 without DMI) can be a functional element to significantly increase the amplitude of the SW passing into the second ferromagnet and the amplitude of the reflected SW compared to the amplitude of the incident spin wave, i.e., can serve as an amplifier for spin waves.

The analytical model of the SW propagation constructed in this work assumes the existence of 2 types of non-reciprocity effects which is a consequence of the existence of 2 methods of manufacturing non-equivalent two-layer systems (the second system becomes the first when rotating the first FM with DMI on 180º during manufacture). There is a possibility of manufacturing 4 non-equivalent double-layer systems of type FM with DMI / FM with DMI at different non-zero values of the constant DMI. That allows to receive 4 types of effects of nonreciprocity in such systems. The nonreciprocity factor increases monotonically with increasing DMI constant in the system FM with DMI / FM without DMI and the system FM without DMI / FM with DMI. The results of calculations demonstrate the possibility to construct the two-layer ferromagnetic systems with high nonreciprocity factors (>10) [42]. The maximum difference of the SW transmission coefficients $\Delta a$ and $\Delta \tilde{a}$ is achieved at large values of the homogeneous exchange constant at the interface $A = 10^{-2}$ $cm$. At large values of the homogeneous exchange constant ($A = 10^{-2}$ $cm$) the maximum difference of the SW reflection coefficients $\Delta r$ and $\Delta \tilde{r}$ is small (for most materials less than 1%, for YIG about 6%). Large maximum difference of the SW reflection coefficients $\Delta r$ and $\Delta \tilde{r}$ is achieved at small values of the homogeneous exchange constant at the interface ($A = 10^{-5} - 10^{-6}$ $cm$) and for example, for YIG at $D = 0.1$ $Erg/cm^2$ reaches more, than 2 times. Thus, double-layer systems such as FM with DMI / FM with DMI at different non-zero values of the constant DMI are promising candidates for logic device applications [42] due to the effect spin wave nonreciprocity.

The boundary conditions at the interface between two ferromagnets with different DMI and results of analytical modelling of SW propagation through the double layer ferromagnet with/without DMI in the first/second layer contributes to the construction of the magnonic crystals with nonreciprocal SW properties due to DMI. Besides, the results of this investigation deepen the knowledge about the SW propagation control in magnonic devices.

**ORCID iDs**

Oksana Gorobets https://orcid.org/0000-0002-2911-6870

Tiukavkina Iryna https://orcid.org/0000-0002-5003-8635

Gerasimenko Rostislav https://orcid.org/0000-0001-8239-6309